\documentclass[epj]{svjour}
\usepackage{graphics}
\usepackage{amssymb,amsmath}
\usepackage{color}
\usepackage{hyperref}
\usepackage{listings}
\usepackage{amsmath}
\usepackage{graphicx}
\usepackage{caption}
\usepackage{subcaption}
\usepackage{bbold}
\usepackage{multirow}
\usepackage{enumerate}
\usepackage[toc,page]{appendix} 

\usepackage[table,xcdraw]{xcolor}

\hypersetup{
    bookmarks=true,         
    unicode=false,          
    pdftoolbar=true,        
    pdfmenubar=true,        
    pdffitwindow=false,     
    pdfstartview={FitH},    
    pdfauthor={},     
    colorlinks=true,       
    linkcolor=blue,          
    citecolor=red,        
}

\newcommand{\ket}[1]{| #1 \rangle}
\newcommand{\bra}[1]{\langle #1 |}
\newcommand{\braket}[2]{\left \langle #1 \middle| #2 \right \rangle}

\definecolor{dkgreen}{rgb}{0,0.6,0}
\definecolor{gray}{rgb}{0.5,0.5,0.5}
\definecolor{mauve}{rgb}{0.58,0,0.82}

\begin{document}
\title{Control Optimization 
for Parametric Hamiltonians by Pulse Reconstruction}
\author{Piero Luchi\inst{1}\inst{2} \and Francesco Turro\inst{1}\inst{4}  \and Sofia Quaglioni\inst{3} \and Xian Wu \inst{3}  \and Valentina Amitrano\inst{1}\inst{2} \and Kyle Wendt\inst{3}  \and Jonathan L Dubois\inst{3}  \and Francesco Pederiva\inst{1}\inst{2}
}                     
%
%
\institute{Physics Department, University of Trento, Via Sommarive 14, I-38123 Trento, Italy \and INFN-TIFPA Trento Institute of Fundamental Physics and Applications, Via Sommarive, 14, I-38123 Trento, Italy \and Lawrence Livermore National Laboratory, P.O. Box 808, L-414, Livermore, California 94551, USA \and Institute of Nuclear Theory (INT), University of Washington,
Physics-Astronomy Building, Box 351550, Seattle, WA 98195-1550}
\date{Received: date / Revised version: date}
%
\abstract{
Optimal control techniques provide a means
to tailor the control pulses required to generate customized quantum gates, which helps to improve the resilience of quantum simulations to gate errors and device noise. However, the significant amount of (classical) computation required to generate customized gates can quickly undermine the effectiveness of this approach, especially when pulse optimization needs to be iterated. We propose a method to reduce the computational time required to generate the control pulse for a Hamiltonian that is parametrically dependent on a time-varying quantity. We use simple interpolation schemes to accurately reconstruct the control pulses from a set of pulses obtained in advance for a discrete set of predetermined parameter values.  
We obtain a reconstruction with very high fidelity and a significant reduction in computational effort. We report the results of the application of the proposed method to device-level quantum simulations of the unitary (real) time evolution of two interacting neutrons based on superconducting qubits.   
%
} 
\maketitle

\section{Introduction}\label{sec:1}
The standard quantum computing (QC) approach is based on expressing arbitrary unitary operations in terms of a set of universal (or primitive) quantum gates, making use of the Solovay-Kitaev theorem \cite{SolovayKitaev}. This gate-based approach has been demonstrated to be, in principle, efficient for the simulation  of complex systems on a quantum computer \cite{divincenzo2000criteria,barends2015superconducting}.
In practice, the performance and reliability of the generated real-time evolution suffer from gate error rates and quantum device noise. In fact, the decomposition of the unitary operator can consist of a large number of gates, resulting in a deep circuit where the sum of individual quantum gate errors degrades the result.  
In addition to extending the coherence time by improving the fabrication process, the design of available qubits~\cite{place2020new,nersisyan2019manufacturing,nguyen2019high} and the readout accuracy \cite{martinez2020improving,luchi2022enhancing}, {another way to improve the noise resilience of quantum simulations is to design efficient quantum control protocols that allow the implementation of arbitrary quantum gates}. This control-based approach tailors a microwave control pulse so that, when applied to the qubits or subsets of them, it realizes the desired unitary transformation in a minimum number of applications and time \cite{palao2002quantum,atia2014quantum,holland2019optimal,chow2010control_qubit,wu2020high}. This approach gives both a shallower quantum circuit and a reduction in the noise produced by deep circuits.  \newline 
\indent The mathematical framework in which we operate is quantum optimal control~\cite{werschnik2007QOC,kirchhoff2018optimized,machnes2011comparing}, defined as the procedure of designing microwave pulses that perform arbitrary unitary transformations using non-linear optimization techniques. The quantum optimal control approach is very useful for simulating quantum mechanical systems \cite{o2016scalable} where the standard gate-based approach performs poorly due to the need for very deep quantum circuits. This approach has been shown to be a promising way to simulate few-nucleon dynamics in a realistic setup \cite{holland2019optimal}. \newline
\indent However, the control-based approach is not perfect. The optimization process requires a considerable amount of classical computing power and time, which grows rapidly with the dimensionality and complexity of the simulated system.
Furthermore, if the desired unitary operator depends on some external or internal time-dependent parameters, the control signal must be recomputed at each simulation time step. This drawback can be effectively addressed by finding a way to parametrize and reconstruct the control pulses, thereby avoiding a full re-evaluation for each realization of the time-dependent parameters. \newline \indent In this paper, we propose a method to reconstruct such control pulses from a limited set of them, computed in advance from a discrete set of points in the Hamiltonian parameter space, using a simple interpolation procedure. 
The method is tested at the device simulation level using as a testbed a set of superconducting qudits \cite{holland2019optimal,wu2020high} 
given by transmon qudits \cite{paik20113Dtrasmon,koch2007transmon}.
In this general configuration, the method is applied to a transverse-field closed-loop Ising model to test the performance and scaling of the method. It is then applied to a simple but realistic case of two neutron dynamics \cite{holland2019optimal} to show how it can help to improve simulation performance. 
This reconstruction approach has been shown to be able to handle control pulses with a wide range of frequency components, for Hamiltonians of different dimensions and with different numbers of parameters. 
Moreover, the method is capable of obtaining high-fidelity optimal control pulses with a reduced computational cost. \newline
\indent The use of a general superconducting multi-level system to test the method provides an indication of how the proposed method is easily applicable and scalable to systems with arbitrary numbers of qubits and controls. Furthermore, in the case of the neutron simulation given in the last section, we assume to work with a single 4-level qudit. This is intended to show how the control-based approach and the proposed method are easily and naturally applicable to qudits, in contrast to the gate-based paradigm. This type of hardware is convenient for quantum simulation and could provide a promising platform for practical computation. \cite{goerz2017charting,neeley2009emulation}. \newline \indent
The optimal microwave control pulses that drive the qubits system were obtained using an optimization algorithm known as Gradient Ascent Pulse Engineering \newline (GRAPE) \cite{khaneja2005Grape2,rowland2012Grape1}. However, quantum optimal control techniques are a very general tool that can in principle be applied to any hardware. It is only necessary to take into account the specific characteristics of the hardware when setting up the optimization procedure. \newline \indent The structure of the present work is as follows.
A brief description of the superconducting circuit adopted
and the quantum-optimal control scheme  
can be found in Sec.~\ref{sec:qubits_and_optimization}. The presentation of the control pulse reconstruction method is given in Sec.~\ref{sec:CPR}, where an analysis of the scaling of the method is also reported.
The application of the proposed method to a neutron-neutron interaction dynamic is reported in Sec.~\ref{sec:application}. Finally, conclusions are drawn in Sec.~\ref{sec:conclusion}.

\section{Device Hamiltonian and Optimization Procedure}\label{sec:qubits_and_optimization}

In the context of quantum optimal control, the state of the quantum processor $\ket{\psi}$ can be described by the usual Schr\"{o}dinger equation:
\begin{eqnarray}\label{eq:general_controlled_schrodinger}
|\dot{\psi}(t)\rangle=-i\left ( H_0 + \sum_{k=1}^{N_{ctrl}}\epsilon_k(t)H_k \right ) \ket{\psi(t)}
\end{eqnarray}
where $\hbar=1$. $H_0$ is the general Hamiltonian of the quantum processor. It depends on the physical properties of the qubits, the qubit couplings and the connectivity between the qubits. $H_k$ matrices are the drive Hamiltonians describing the external interaction with the system. These are also device-specific. Finally, $\epsilon_k(t)$ are $N_{ctrl}$ functions describing how the strength of each $H_k$ on the system changes as a function of time. The number $N_{ctrl}$ of such controls depends on the specific hardware and the connectivity between the qubits. The optimal control problem consists in finding the best control shapes $\epsilon_k(t)$ that, when applied to the quantum device for a total time $\tau$, drive the system to the desired state $\ket{\psi(t=\tau)}$ within an acceptable error.  \newline \indent 
We can solve the Schr\"{o}dinger equation \eqref{eq:general_controlled_schrodinger} and obtain:
\begin{eqnarray}\label{eq:general_controlled_schrodinger_solution}
\ket{\psi(t)}=\exp\left \{-i\int_0^t\left ( H_0 + \sum_{k=1}^{N_{ctrl}}\epsilon_k(t)H_k \right )dt \right \} \ket{\psi(0)}
\end{eqnarray}
where the exponential is the unitary matrix $U_0(t)$ i.e. the propagator of the system. It transforms the initial qubit state into the final one.

Now suppose we have an arbitrary Hamiltonian $H_{syst}$ of a quantum system. The dynamics of such a system can be described by another Schr\"{o}dinger equation, $|\dot{\phi}(s) \rangle=-i H_{syst}\ket{\phi(s)}$, where $\ket{\phi(s)}$ is the state of the system at time $s$. Its propagator, $U_{syst}(s)$, can be written:
\begin{eqnarray}\label{eq:general_system_schrodinger_solution}
 U_{syst}(s)=e^{-isH_{syst}}.
\end{eqnarray}
Note that the time $t$ of Eq. \eqref{eq:general_controlled_schrodinger_solution}
is the real time of the experimental device, while the time $s$ of Eq. \eqref{eq:general_system_schrodinger_solution} is the time of the quantum system itself.

If we now want to simulate the dynamics of the quantum system with the quantum processor, we compute the short-time propagator $U_{syst}(\delta s)$, which describes how the physical system changes in a small time step $\delta s$, and we optimize the controls $\epsilon_k(t)$ so that the device propagator $U_0(\tau)$, for an appropriate control duration $\tau$, transforms the qubit state in the same way as $U_{syst}(\delta s)$ would. Having previously defined a mapping between the two systems, we are able to infer the dynamics of the quantum system from that of the quantum processor.

In practice, this is achieved by many optimization algorithms. In this paper, we rely on the GRAPE algorithm \cite{rowland2012Grape1,khaneja2005Grape2} implemented in the Qutip Python package \cite{qutip}. It works by optimizing the controllers $\epsilon_k(t)$ such that, within a given accuracy threshold, they satisfy the equality between $U_{syst}(\delta s)$ and $U_0(\tau)$, i.e:
\begin{eqnarray}\label{eq:GRAPE_general}
U_{syst}(\delta s)=\mathcal{T} \exp\left \{ -i \int^{\tau}_0  \left ( H_0 + \sum_{k=1}^{N_{ctrl}}\epsilon_k(t)H_k \right ) dt \right \}
\end{eqnarray}
where $\mathcal{T}\exp{}$ stands for time-ordered exponential and $\tau $ is the total duration of the $N_{ctrl}$ control pulses $\epsilon_k(t)$. The quantity which is optimized by GRAPE is the fidelity $\mathcal{F}$:
\begin{eqnarray}\label{eq:fidelity}
\mathcal{F}=1-\tilde{\mathcal{D}},
\end{eqnarray}
where $\tilde{\mathcal{D}}$ is the normalized Hilbert-Schmidt norm,
\begin{eqnarray}
\tilde{\mathcal{D}}=\frac{1}{2} - \frac{1}{2d }\ \mathrm{Re} \ \mathrm{Tr}(U_{syst}^\dagger (\delta s)U_0(\tau)),
\end{eqnarray}
with $d=\dim(U_{syst}(\delta s))$ the dimension of the matrices. This metric quantifies the similarity of the two matrices by giving a value close to one when they are similar and close to zero when they are different.
\\
In this work, we assume to work with {qudits (i.e. a generalization of the qubit with more than two levels) realized by transmon superconducting devices} in the dispersive regime in the limit of low nonlinearity $\alpha$ \cite{place2020new,koch2007transmon,blais2004cavity}. The general Hamiltonian $H_0$ for a set of $N_q$  transmon qudits can be described by an appropriately parameterized Jaynes-Cumming model \cite{blais2020circuit,kairys2021efficient,nigg2012Transmon_Hamilt}:
\begin{eqnarray}\label{eq:device_Ham}
H_0=&&\sum_{i=1}^{N_q} \left ( \omega_i b_i^{\dagger}b_i + \frac{\alpha_i}{2}b_i^{\dagger}b_i(b_i^{\dagger}b_i-1) \right ) + \\
&&+\sum_{i,j=1}^{N_q}g_{i,j}( b_i^{\dagger}b_j \nonumber +b_ib_j^{\dagger}),
\end{eqnarray}
where the $b_i^{\dagger}(b_i)$ is the creation (annihilation) operator of $i^{th}$ qudit, $\omega_i$, $\alpha_i$ are the $i^{th}$ qudit frequency and anarmonicity and $g_{i,j}$ are the coupling constant between qudit $i$ and qudit $j$. In general, the first term represents the sum of individual qudit Hamiltonians and the second term represents the coupling between qudits. Different qudit connectivity can be simulated by modifying  $g_{i,j}$ values. In general, the system could be coupled to a cavity/resonator in order to perform qudits measurements \cite{blais2004cavity,krantz2019superconducting_qubits,Blais2021cQED,kirchhoff2018optimized}. However, the qudits are well detuned from the cavity and the all-microwave control is
generally performed with direct drives, so we can neglect the
dynamics of the cavity in the optimization, taking also into account that {$\langle b_i^\dagger b_i\rangle\approx 0$}  \cite{Blais2007QED,wu2020high}.

The individual control Hamiltonians are $H_k=(b_k^{\dagger}+b_k)$, where the index $k$ indicates the qudit to which they are applied. The whole control Hamiltonian $H_c(t)$ reads:
\begin{eqnarray}
H_{c}(t)=\sum_{k=\mathcal{I}_c} \epsilon_k(t)(b_k^{\dagger}+b_k),
\end{eqnarray}
where $\mathcal{I}_c$ is the set of controlled qudit.  

The discrete version of GRAPE equivalence Eq. \eqref{eq:GRAPE_general} becomes:
\begin{eqnarray}\label{eq:GRAPE}
U_{syst}(\delta t)= \mathcal{T} \exp\left \{ -i \delta \tau \sum_{t_i=0}^{N_{\tau}} \left ( H_0 + H_c(t_i) \right ) \right \}, 
\end{eqnarray}

\noindent where $N_{\tau} $ is the number of time steps of duration $\delta \tau$ into which the pulse of total duration $\tau$ is divided.

\section{Control Pulse Reconstruction (CPR) method} \label{sec:CPR}

The main drawback of using optimization algorithms (such as GRAPE) to compute optimized control sequences $\epsilon_k(t)$ is the large computational cost. In fact, the time needed to compute a control grows exponentially with the number of qudits involved (i.e. the dimension of the matrix $U_{syst}(\delta s)$). Furthermore, the control impulses are not transferable, i.e. any change in the unitary operator requires a completely new optimization. This is especially true if there is any parametric dependence of the Hamiltonian $H_{syst}$ and a study of the system as a function of the parameter value is required. An example, which will be shown in detail in Sec. \ref{sec:application}, is the simulation of a scattering process of two nucleons interacting through a spin/isospin dependent interaction, as captured by the leading order (LO) in the chiral effective field theory expansion. As described in Ref. \cite{holland2019optimal} for the case of two neutrons, it is possible to decompose the propagator into a spin-independent and a spin-dependent part, the latter being parametrically dependent on the instantaneous specific value of the coordinates of the neutrons. This implies that the control pulses implementing the instantaneous spin dynamics have to be computed at each time step of the neutron's physical trajectory. This tends to neutralize the benefits in terms of computational speed-up that quantum computing brings to the simulation of quantum processes.
\par
An explicit calculation of the pulses for each value of the Hamiltonian parameters can be avoided by the following general Control Pulse Reconstruction (CPR) method.
Given the Hamiltonian $H_{syst}(\lambda)$, where $\Lambda=$ \newline
${[\lambda_1,\lambda_2,\dots,\lambda_K]}$ is a point in the $K$-dimensional space of parameters characterizing the Hamiltonian, and assuming that one wants to implement the unitary transformation $U_{syst}(\delta s,\Lambda)=\exp \left\{-i\delta s H_{syst}( \Lambda)\right \} $ on the quantum device, one should: 
\begin{enumerate}
    \item Solve Eq. \eqref{eq:GRAPE} (using GRAPE or another equivalent optimization algorithm) for a discrete grid of values of $\Lambda$ and store the resulting control pulses. The result is a family of controls, each of which implements the specific unitary transformation for the corresponding value of $\Lambda$.
    
    \item Perform a fit of the pulses either in terms of a function (e.g. a polynomial fitting) or as an expansion over a basis (e.g. Fourier transform of the pulses). Let us call $C=[c_1,c_2,\dots,c_M]$ the $M$ coefficients of the fit (e.g. the polynomial coefficient or some type of coefficient relative to the expansion over a basis). 
    
    \item Find a mathematical relationship between the coefficients of the fit $C$ and the parameters $\Lambda$, i.e. $C=f(\Lambda)=[f_1(\Lambda),f_2(\Lambda),\dots,f_M(\Lambda)]$ 
    
    \item Reconstruct the pulses for an arbitrary $\tilde{\Lambda}$ via the following procedure: select the $\tilde{\Lambda}$ values of interest, recover the fit parameters $\tilde{C}$ through the mathematical relationship $C=f(\Lambda)$ and reconstruct, finally, the control pulses identified by $\tilde{C}$ inverting the fit relation.
    
\end{enumerate}
The procedure of step 4 is used in the simulation loop to recover all the needed control pulses.

\subsection{CPR Method Realization}\label{sec:CPR_realization}
In this section, we describe in detail the actual implementation of the CPR method we use in this work.

First of all, let's define the grid of $ \Lambda$ values. Each component $\lambda_i $ of $ \Lambda$, with $i=1,...,K$, is a continuous parameter of the Hamiltonian $H_{syst}(\Lambda)$ (and the corresponding propagator $U_{syst}(\Lambda)$) defined on a real interval $I_i$. Let's divide each $I_i$ in $Z_i$ discrete values, obtaining $Z_i$ discrete values of $\lambda_i$, namely $\lambda_i^{a_i}$ for $a_i=1,...,Z_i$. 
Let's define $\Lambda_m$ to be a K-dimensional grid of all possible combination of these parameters, i.e $\Lambda_m^{a_1,a_2,...,a_K}=[\lambda_1^{a_1}, \lambda_2^{a_2}...,\lambda_K^{a_k}]$ with $a_1\in \{1,2,...,Z_1\}$, $\ a_2\in \{1,2,...,Z_2\}$, $..., \ a_K\in \{1,2,...,Z_K\}$. $\Lambda_m$ is the discrete grid of value of the parameters $\Lambda$.
  
Second, we define the expansion $C$. The pulses have shapes that cannot be easily interpreted in terms of elementary functions, since they may contain multiple frequency components. This is especially true for large Hamiltonians. To deal with this type of signal, an obvious choice of expansion is to use the Fourier transform. In this case, the fit parameters $C$ are (an appropriate subset of) the components of the control pulse spectra.

Finally, the mathematical relation $C=f(\Lambda)$ is realized by a multi-linear interpolation of each component of the vector $C$ over the grid $\Lambda_m$.  

\noindent To summarize, in this work the CPR method takes the following form:
\begin{enumerate}

     \item For each element $\Lambda' \in \Lambda_m$, corresponding to a single and unique combination $[\lambda_1^{a_1}, \lambda_2^{a_2}...,\lambda_K^{a_k}]$, we can compute, via Eq. \eqref{eq:GRAPE}, the $N_{ctrl}$ controls $\epsilon_z(t,\Lambda')$, with $z=1,...,N_{ctrl}$, implementing the specific transformation identified by $U_{syst}(\Lambda')$. Let $\mathcal{E}_z(t,\Lambda_m)$ be the discrete set of all $\epsilon_z(t,\Lambda')$  for all $\Lambda' \in \Lambda_m$.
    
    \item Let $C_z(\omega,\Lambda')=[c_{z,1}(\omega_1,\Lambda'),..,c_{z,M}(\omega_M,\Lambda')]$ be the (discrete) Fourier transform of  $\epsilon_z(t,\Lambda')$ truncated to its $M^{th}$ component. $C_z(\omega,\Lambda')$ is interpreted as the expansion coefficients vector over the basis (i.e. the frequencies components $\omega_i$). Obviously, as a result of a Fourier transform, each $C_z$ is a complex vector $C_z=C_{z}^{real}+iC_z^{imag}$. Let $\mathcal{C}_z(\omega,\Lambda_m)$ be the discrete set of all $C_z(\omega,\Lambda')$ for all $\Lambda'\in \Lambda_m$.  
    
    \item To obtain a new vector $\tilde{C}_z(\omega,\tilde{\Lambda})$ for an arbitrary $\tilde{\Lambda} \notin \Lambda_m$, we use a linear multivariate interpolation of each component $c_{z,i}$ between the values of the subset\newline $\mathcal{C}_z(\omega_i,\Lambda_m)$ over the K-dimensional grid $\Lambda_m$.  This defines the mathematical relation $C=f(\Lambda)$ that links the parameters $\Lambda$ to the expansion coefficients $C_z$. The interpolation is performed separately for the real and imaginary parts of the vector $\tilde{C}_z(\omega,\tilde{\Lambda})$. This is done for each $z =1,..,N_{ctrl}$.
    
    \item The control pulse reconstruction procedure becomes the following: the $N_{ctrl}$ control pulses for an arbitrary $\tilde{\Lambda} \notin \Lambda_m$ are recovered by interpolating, for each control $z$, the real and imaginary part of the new coefficients $\tilde{C}_z(\tilde{\Lambda})$ through the linear multivariate interpolation of the set $\mathcal{C}_z(\omega_i,\Lambda_m)$ for $i=1,...,M$,  over the discrete  K-dimensional grid $\Lambda_m$. The resulting $\tilde{C}$ is then transformed back into the time domain using the inverse Fourier transform to obtain the corresponding new controls $\tilde{\epsilon}_z(t,\tilde{\Lambda})$, which are not elements of the original sets $\mathcal{E}_k(t,\Lambda_m)$.

\end{enumerate}

\begin{figure*}[h]
    \centering
    \includegraphics{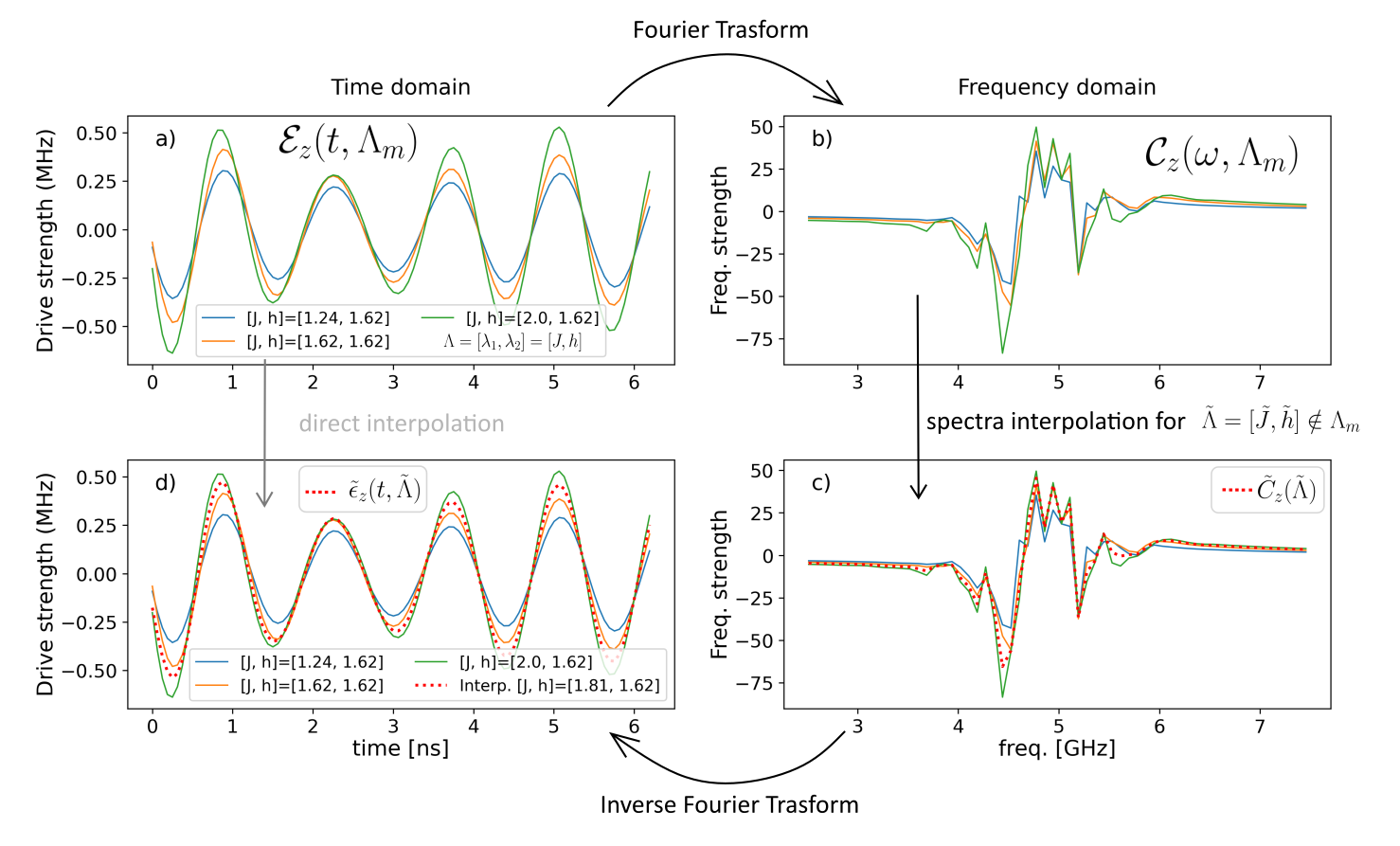}
    \caption{A graphic description of the CPR method. Panel (a) shows a subset of $\mathcal{Epsilon}_z(t,\Lambda_m$ controls for soome values of $\Lambda=[\lambda_1, \lambda_2]$ (which in this case are the $J$ and $h$ parameters of the Ising model described in Sec.\ref{sec:CPR}). By Fourier transforming each control, the set of corresponding spectra $\mathcal{C}_z(\omega,\Lambda_m)$ is obtained, represented in panel (b). In this case, only the real part of this set is represented, for the sake of simplicity. By taking a new value $\tilde{\Lambda}$ of the parameters, a new spectrum $\tilde{C}_z(\omega,\tilde{\Lambda})$ can be interpolated, represented with the dashed red line in panel (c). As can be seen, the new spectrum lies between the others. Transforming the new spectrum with the inverse Fourier transform, as in panel (d), the new control $\tilde{\epsilon}(t,\tilde{epsilon})$ for the new Lambda parameters can be finally obtained. As in the case of the spectrum, the new control falls between those calculated with the parameter values just above or below. One could also switch from panel (a) to panel (d) with a direct interpolation of the controls.}
    \label{fig:CPR_method_graphical_explanation}
\end{figure*}

\noindent A graphical representation of the working principle of the CPR method is shown in Fig. \ref{fig:CPR_method_graphical_explanation}. The diagram refers to the controls obtained for the Ising model presented in the following section.

{One might wonder why one would expand the signal using the Fourier transform rather than interpolating directly between each time step of the signals as shown in Fig. \ref{fig:CPR_method_graphical_explanation}(panels a and d), and why one would choose the Fourier transform.}

The answer to the first question is that the number of spectral components relevant to the reconstruction of a control is much smaller than the number of time steps of the control itself. In fact, as can be seen in Fig. \ref{fig:CPR_method_graphical_explanation}, while the controls (panel a) all have values different from zero, many spectral components (panel b) are close to zero, so the expansion over the Fourier basis offers a significant reduction in the number of elements that one needs to interpolate using the CPR method. In practice, in this work, all spectra were truncated to the $M^{th}$ component, after which the values of the spectra fall below a threshold. This threshold was chosen to be the maximum value between 0.001 and the higher value of the spectra divided by 1000. This allowed the $M$ to be two or three orders of magnitude smaller than the number of time steps of the controls (depending on the configuration). In Sec. \ref{sec:CPR_direct_vs_fft}, analyses supporting these claims are given.

The answer to the second question is simply that the Fourier transform is the most straightforward transform to apply to time signals. Moreover, its implementation in coding languages is very efficient thanks to "fast Fourier transform" algorithms. In addition, the Fourier transform offers the possibility of using the CPR method with noisy controls by acting as a high-frequency filter. This could be the case for certain types of optimization initial conditions not used in this work.

\subsection{CPR Method Characterization}

\subsubsection{Scaling}

To test the scaling of the method, we investigate the fidelity of the interpolated controls in three cases: 1) the number of quantum levels/qudits involved, 2) the number $K$ of Hamiltonian parameters, and 3) the $\Lambda_m$ grid spacing. 
This is tested on a 1D transverse-field Ising model with periodic boundary conditions (i.e. in a closed spin chain configuration). We have chosen the Ising model because it is a simple and well-known model, easy to map onto the quantum processor, but, at the same time, it is a non-trivial system.
The transverse-field Ising Hamiltonian is:

\begin{eqnarray}\label{eq:ising_hamiltonian}
H_{Ising}=\sum_{i=1}^N J_i\sigma^z_i\otimes\sigma^z_{i+1}
+h\sum_{i=1}^N \sigma^x,
\end{eqnarray}
\noindent with $N$ the number of spins/sites of the chain, $\sigma^z_i$ and $\sigma^x_i$ the $z$ and $x$ Pauli matrices for the $i^{th}$ spin/site, $J_i$ the coupling terms between nearest spins and $h$ the external field intensity. Note that we set $N+1\equiv 1$ to obtain the closed chain. The propagator is written:
\begin{eqnarray}\label{eq:ising_prop}
U_{Ising}=e^{-i H_{Ising}\delta s}.
\end{eqnarray}
for an arbitrary (small) time interval $\delta s$. \newline  \indent Each spin has two states, the spin up, $\ket{\uparrow}$, and the spin down, $\ket{\downarrow}$. The mapping of the Ising model in the quantum processor is therefore simply done by identifying each spin with a single two-level qudit of the quantum processor Hamiltonian of Eq. \eqref{eq:device_Ham}.
Using Eq. \eqref{eq:GRAPE}, it is now possible to calculate the controls $\epsilon_z(t)$ that implement $U_{Ising}$ in the quantum device. For this experiment, we use one control for each qudit (but many other configurations can be realized). The parameters of the device Hamiltonian are given in Tab.\ref{tab:qudit_param}.

\begin{table}[]
\begin{tabular}{llllll}
\multicolumn{1}{l|}{} & \multicolumn{1}{l|}{Q 1} & \multicolumn{1}{l|}{Q 2} & \multicolumn{1}{l|}{Q 3} & \multicolumn{1}{l|}{Q 4} & \multicolumn{1}{l|}{Q 5} \\ \hline
\multicolumn{1}{l|}{$\omega / 2 \pi$ [GHz]} & \multicolumn{1}{l|}{5.114} & \multicolumn{1}{l|}{4.914} & \multicolumn{1}{l|}{4.714} & \multicolumn{1}{l|}{4.614} & \multicolumn{1}{l|}{4.514} \\ \hline
\multicolumn{1}{l|}{$\alpha / 2 \pi$ [GHz]} & \multicolumn{1}{l|}{-0.33} & \multicolumn{1}{l|}{-0.33} & \multicolumn{1}{l|}{-0.33} & \multicolumn{1}{l|}{-0.33} & \multicolumn{1}{l|}{0.33} \\ \hline
\multicolumn{1}{l|}{$g_{i,j}/ 2 \pi$ [GHz]} & \multicolumn{1}{l|}{0.0038} & \multicolumn{1}{l|}{0.0038} & \multicolumn{1}{l|}{0.0038} & \multicolumn{1}{l|}{0.0038} & \multicolumn{1}{l|}{0.0038}
\end{tabular}
\caption{Values of qudit parameters for Eq. \ref{eq:device_Ham} used in the Ising model analysis.}
\label{tab:qudit_param}
\end{table}

We consider the parameters $J_i \ \forall i\in[1,N]$ and $h$ all taking values in the same real interval $I=[0.2,2]$. We discretize $I$ in $Z$ discrete values so to have the discrete values $J_i^{a_i}$ and $h^b$ with $a_i,b\in\{1,..,Z\}$. We can now define $\Lambda_m$ as the $N+1$-dimensional grid of all possible combination of $J_i^{a_i}$ and $h^b$, i.e.  $\Lambda_m^{a_1,...,a_N,b}=[J_1^{a_1},..,J_N^{a_N},h^b]$. We can now compute the $N_{ctrl}$ sets of controls, $\mathcal{E}_z(t,\Lambda_m)$, and the corresponding set of truncated spectra, $\mathcal{C}_z(\omega,\Lambda_m)$, for $z\in\{1,..,N_{ctrl} \}$. Now we are finally able to obtain new controls following the CPR procedure described above. See, again, Fig. \ref{fig:CPR_method_graphical_explanation} for a graphical representation of the method.

\paragraph{Fidelity vs. number of qudits}

\begin{figure}[h]
\includegraphics[width=\columnwidth]{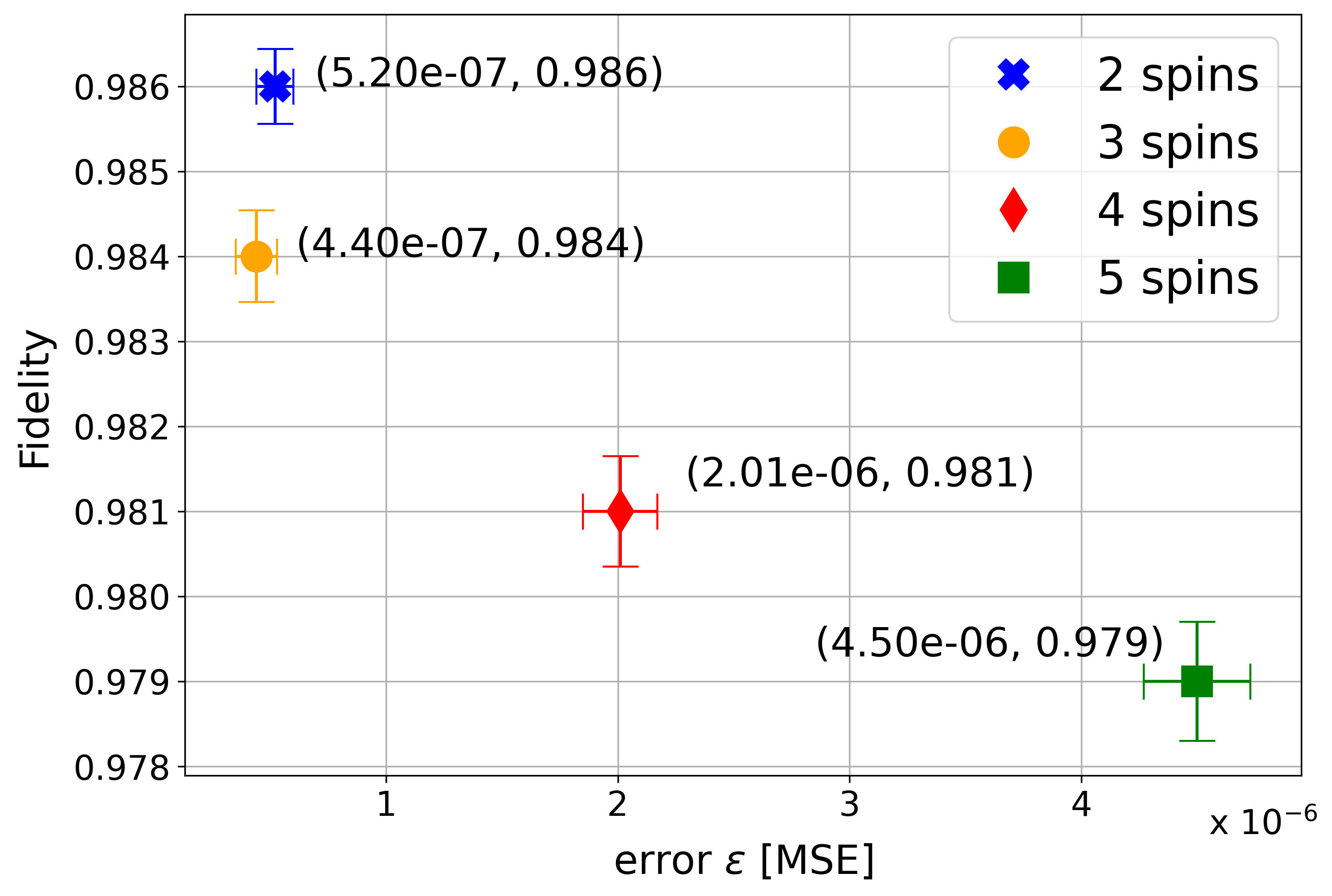}
\caption{The x-axis represents the average mean squared error (MSE) between the controls optimized with GRAPE and the controls interpolated with the CPR method for the same sampled set of parameters $h$ and $J$. The y-axis instead represents the average fidelity between the exact propagator $U_{Ising}$ and the reconstructed one, $\tilde{U}_{Ising}$, with the right-hand side of eq. \eqref{eq:GRAPE}. Each point refers to a chain with a different spin number $N$.}
\label{fig:fidelityvserror}
\end{figure}

As the number of spins $N$ increases, so does the dimension of the matrix $U_{Ising}$. The optimization becomes more difficult and the control pulses contain higher and higher frequencies. Therefore, it is necessary to study how the CPR method scales with the dimension of the system. 

We consider Ising models with increasing number $N$ of spins with two parameters, the values $h$ and $J_i=J$ for each $i=1,..., N$. The number $Z$ of subintervals of $I$ is set to 9. So we have a 2-dimensional grid $\Lambda_m^{a,b}=[J^{a},h^b]$ with $a,b=1,...,9$. We prepare the CPR method dataset by calculating $\mathcal{E}^N_{z}(t,\Lambda_m)$ for $N=2,...,5$. The average fidelity of the datasets computed with GRAPE is $0.981 \pm 1.4 e-4$ for all $N$.
The scaling is tested with the following analysis. We sample new values $\tilde{J}$ and $\tilde{h}$ from a uniform probability distribution in the interval $I$ and, based on these values, we compute the controls using both the CPR method, $\tilde{\epsilon}_z^N(t,[\tilde{J},\tilde{h}])$, and the GRAPE algorithm, $\epsilon_z^N(t,[\tilde{J},\tilde{h}])$. We then compute the mean square error (MSE), $\varepsilon$, between the two controls. We also reconstruct the propagator, $\tilde{U}_{Ising}([\tilde{J},\tilde{h}])$, induced by the interpolated controls, $\tilde{\epsilon}_z^N(t,[\tilde{J},\tilde{h}])$, by inserting it into the right-hand side of Eq. \eqref{eq:GRAPE}. In this way we can calculate the fidelity (Eq.\eqref{eq:fidelity}) between the exact propagator $U_{Ising}([\tilde{J},\tilde{h}])$ defined by Eq. \eqref{eq:ising_prop} and the one induced by the controls, i.e.  $\tilde{U}_{Ising}([\tilde{J},\tilde{h}])$, found with Eq. \eqref{eq:GRAPE}, to check their degree of similarity. This procedure is repeated 50 times for each $N$ and the results are averaged. Fig. \ref{fig:fidelityvserror} plots the reconstruction fidelity against the MSE. As expected, the reconstruction fidelity decreases on average as the number of spins increases, while the error increases. However, the MSE remains very limited in absolute value and the decrease in fidelity remains small compared to the dataset's average fidelity of $0.981$. Furthermore, the decreasing trend suggests that the fidelity degradation slows down as the number of spins increases.

\paragraph{Fidelity vs. number of Hamiltonian's parameters}
The Hamiltonian can depend on several $\lambda_i$ parameters, so it is interesting to study the scaling of the CPR method with respect to this number. We study its behavior on a 3-spin chain, taking as $\lambda_i$ the parameters $J_1, J_2, J_3$ and $h$. Each of them varies in the interval $I$ divided into $Z=9$ discrete values. We compute the first datasets $\mathcal{E}^1_k(t,\Lambda_m^1)$ varying only $h$ and fixing $J_1, J_2, J_3$ to 1, the second datasets $\mathcal{E}^2_k(t,\Lambda_m^2)$ varying only $h$ and $J_1$ while fixing $J_2, J_3$ to 1, and so on until all four parameters are used. The time to compute the dataset grows (exponentially) with the number of parameters since the number of controls in a dataset is $\prod_i^{n_{par}}Z$, where $n_{par}$ is the number of parameters used. The fidelity of the controls in the dataset is 0.981 for each $n_{par}$. Then, as in the previous paragraph, we take new random values $\tilde{h},\tilde{J_1},\tilde{J_2}, \tilde{J_3}$ and we interpolate the new controls $\tilde{\epsilon}_z^{n_{par}}(t)$ with the CPR method for increasing $n_{par}$. We then reconstruct the propagator $\tilde{U}_{Ising}$ induced by these controls via Eq.\eqref{eq:GRAPE} and we compute its fidelity with respect to the exact one obtained via Eq.\eqref{eq:ising_prop} for the same sampled parameter values. The process is performed 50 times for each $n_{par}=1,2,3,4$ and the results are averaged. Tab. \ref{tab:fidelity_vs_num_par} reports the results. The first column reports the average fidelity of the controls in the datasets $\mathcal{E}^{n_{par}}_k(t,\Lambda_m^{n_{par}})$. In the second column, the average fidelity of the control-induced propagators for random sampling of $J_i$ and $h$ is reported. It can be seen that the average fidelity for the interpolated controls remains high and comparable to the fidelity of the dataset (obtained with GRAPE). The data suggest that the CPR method is able to obtain controls with good fidelity for each configuration, showing good behavior for an increasing number of Hamiltonian parameters.

\begin{table}[]
\begin{tabular}{l|l|l}
$n_{par}$ & \begin{tabular}[c]{@{}l@{}}aver. fid. \\ dataset\end{tabular} & \begin{tabular}[c]{@{}l@{}}aver. fid. \\ sampling\end{tabular}  \\ \hline
\rowcolor[HTML]{EFEFEF} 
1 & 0.981 $\pm$ 1.3 e-4 & 
0.981 $\pm$ 2.4 e-4    \\
2 &  0.981 $\pm$ 1.2 e-4 &  0.981 $\pm$ 5.4 e-4  \\
\rowcolor[HTML]{EFEFEF} 
3 & 0.981 $\pm$ 6.4 e-4  &  0.981 $\pm$ 1.2 e-3   \\
4 & 0.981 $\pm$ 4.5 e-5  & 0.982 $\pm$ 2.3 e-4    
\end{tabular}
\caption{Average fidelity for a 3-spin Ising chain for an increasing number of Hamiltonian parameters. First column: average fidelity of the dataset controls. Second column: average fidelity of CPR-interpolated controls with uniformly sampled $J_i$ and $h$. }
\label{tab:fidelity_vs_num_par}
\end{table}

\paragraph{Fidelity vs. density of $\Lambda_m$ grid}
Another test to characterize the CPR method concerns the behavior of the reconstruction fidelity as a function of the discretization density of the interval $I$. We take a 3-spin closed chain with two parameters, $h$ and $J_i=J$ for each $i=1,2,3$, and compute the datasets $\mathcal{E}^Z_{k}(t,\Lambda_m)$ for the interval $I$ divided into $Z=3,5,10,15$ subintervals. The data fidelity was again set to 0.981 for each case. As in the previous case, we sample a new value of the parameters, interpolate a new control and compute the fidelity of the reconstructed propagator with respect to the exact one. This is repeated 50 times for each $Z$ and the results are averaged. In Tab. \ref{tab:fidelity_vs_mesh} the results are reported in the same format as in the previous case. As expected, the fidelity of the CPR method is low for small $Z$ as the linear interpolation becomes imprecise as it is performed over distant mesh points. However, the analysis shows that the fidelity quickly saturates with increasing discretization density $Z$.

\begin{table}[]
\begin{tabular}{l|l|l}
$n_{par}$ & \begin{tabular}[c]{@{}l@{}}aver. fid. \\ dataset\end{tabular} & \begin{tabular}[c]{@{}l@{}}aver. fid. \\ sampling\end{tabular}  \\ \hline
\rowcolor[HTML]{EFEFEF} 
3 & 0.981 $\pm$ 4.7 e-4 & 
0.952 $\pm$ 4.1 e-3   \\
5 &  0.981 $\pm$ 2.7 e-4 &  0.977 $\pm$ 1.9 e-3 \\
\rowcolor[HTML]{EFEFEF} 
10 & 0.982 $\pm$ 1.9 e-4  &  0.983 $\pm$ 3.5 e-4  \\
15 & 0.982 $\pm$ 9.5 e-4  & 0.983 $\pm$ 1.5 e-4  
\end{tabular}
\caption{Average fidelity for a 3-spin chain with two parameters of the Hamiltonian, $J=J_1=J_2=J_3$ and $h$ for increasing parameters interval $I$ spacing $Z$. The data format is the same as the Tab. \ref{tab:fidelity_vs_num_par}.}
\label{tab:fidelity_vs_mesh}
\end{table}

\subsubsection{Computational cost and time}
Careful consideration should be given to the calculation time. In fact, the method requires a certain number of controls to be calculated in advance, which can result in a prohibitively long calculation time. 

\begin{figure}[h]
\includegraphics[width=\columnwidth]{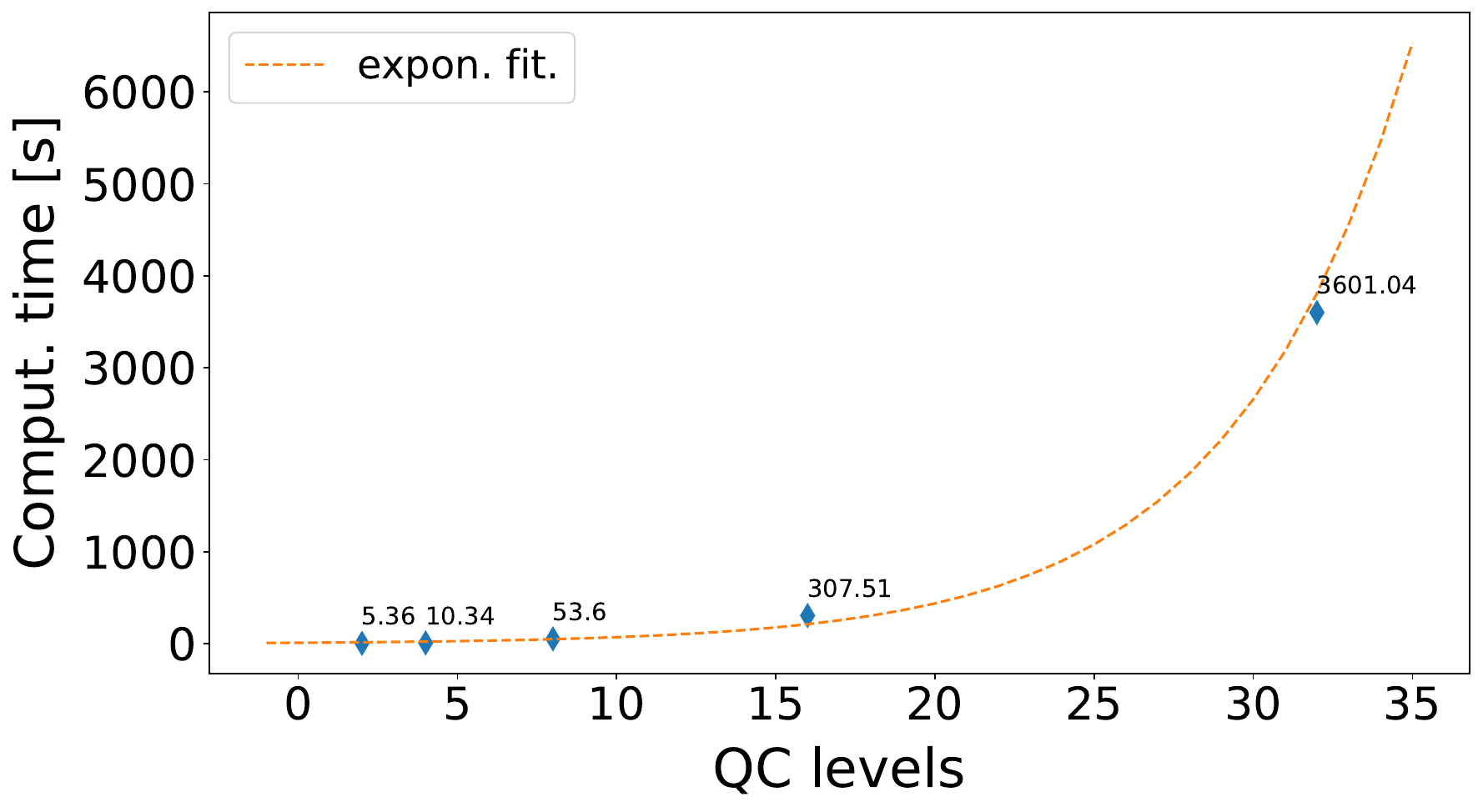}
\caption{Computational time vs. the number of levels of a quantum device. The single time data (diamond markers) are computed for $2^N$ QC levels with $N=[1,2,3,4,5]$ (which identify the total number of levels for 1,2,3,4 or 5 qudits). Computational time grows exponentially with the number of levels, as exponential fitting highlights (dashed line).}
\label{fig:comput_time}
\end{figure}

The average time $t_{G}$ to obtain a control with GRAPE depends exponentially on the number of Hamiltonian dimensions (i.e. quantum device levels or qudits used) and linearly on the number of control time steps (i.e. sampling frequency multiplied by control time duration). As an example, Fig. \ref{fig:comput_time} shows the time $t_G$ to compute 1600 time-step controls for an increasing number of quantum levels of the device. 
Instead, the average time $t_{CPR}$ that the CPR method takes to compute a single control is $0.78$ seconds, regardless of the number of levels/quits, as it is a simple interpolation. These values are based on a mid-range computer with a 2.9 GHz, 4-core processor and 8 GB of RAM.

The total number $A$ of controls in a general controls dataset $\mathcal{E}_z(t,\Lambda_m)$ is:
\begin{eqnarray}
A=\sum_{i=1}^{K}Z_i,
\end{eqnarray}
with $K$ the number of Hamiltonian parameters $\lambda_i$ and $Z_i$ the number of discrete values of interval $I_i$. The total time $T_G$ to compute the $A$ controls with GRAPE is clearly:
\begin{eqnarray}
T_G=t_GA=t_G\sum_{i=1}^{K}Z_i.
\end{eqnarray}

\noindent In order to gain a computational advantage from using the CPR method over using GRAPE, the total number $P$ of controls you want to get from using the CPR method in a simulation must be greater than $A$. Possibly, this difference should be high.

Let's make two examples, assuming we have an advantage when $P>A$. Let $t_G=53.6 s$ and $A=500$. To get an advantage, the time to compute $P$ controls with GRAPE must be greater than the time to compute them with the CPR method plus the time to compute $A$ controls of the dataset in advance. Thus $t_GP>(t_{CPR}P+t_GA)$, from which it is easy to deduce that $P>507$.\newline \indent Conversely, let $P=1000$ be the number of controls used in a simulation. Using GRAPE one would take $t_GP=53.6s \times 1000=53600s \approx 14h$. Otherwise, using the CPR method with $A=500$, one takes $t_GA+t_{CPR}P=53.6s \times 500 + 1000\times 0.78 s =27508 \approx 7.6h$. In fact, the advantage is even greater, because by carefully choosing the initial guess of the GRAPE algorithm and by exploiting parallel computing routines, the time to compute the $A$ controls of the dataset can be significantly reduced. This is generally not possible without using the CPR method, since the controls are computed in series for the subsequent values that the parameters of the Hamiltonian take during the dynamics of the system under analysis.

Therefore, the use of the CPR method requires careful consideration and selection of the appropriate trade-off between dataset size, precision and computational time.

\subsubsection{{Direct and Forurer Trasform interpolation}}\label{sec:CPR_direct_vs_fft}
The CPR method described in Sec. \ref{sec:CPR_realization} works by interpolating between the spectral components of the controls dataset. Of course, one could interpolate directly between the controls to obtain a new control for any value of the $\Lambda$ parameters. However, The complexity of the CPR method presented is justified by the fact that it allows us to interpolate a much smaller number of values by obtaining a small but consistent interpolation speed-up. The procedure described in section \ref{sec:CPR_realization} defines a rigorous method for discarding a set of values from the spectra of the controls based on their relative magnitude. 

Below is a simple example showing the differences between direct interpolation and CPR.

Let's consider the 3-spin Ising model. We use 3 controls, one for each qubit on which the Ising model is mapped. We choose a pulse duration of 75 ns and a sampling rate of 16 GHz or $75 \times 16=1200$ timesteps.

\textbf{Direct interpolation}: As we need to interpolate 3 controls, the total number of values to interpolate is $3 \times 1200=3600$. The average interpolation time for this particular configuration is $1.102 \pm 0.020$ s (over 20 interpolations).

\textbf{CPR method intepolation}: The number of spectral values retained is $M=67$. Consequently, the total number of values to be interpolated is $67 \times 2 \times 3=402$ (the factor $2$ is due to the fact that each spectrum is composed of a real and an imaginary part). In this case the 
average interpolation time is reduced to $0.444 \pm 0.015$ s (over 20 interpolations).

In Tab. \ref{tab:fft_vs_direct} the same calculations are performed for the 2,4 and 5 spin Ising model. These analyses show that the CPR method with Fourier transform offers a speed-up with respect to direct interpolation. The last column of the table shows the ratio between the direct interpolation and the CPR interpolation time. In all cases, the CPR method is about twice as fast. This is a small speed-up, but it can be significant in a context where many interpolations have to be performed.

\begin{table}[]
\begin{tabular}{c|c|c|c|c}
$N$ & $M$ & \begin{tabular}[c]{@{}c@{}}CPR interp.\\  time {[}s{]}\end{tabular} & \begin{tabular}[c]{@{}c@{}}Direct interp. \\ time {[}s{]}\end{tabular} & ratio \\ \hline
\rowcolor[HTML]{EFEFEF} 
2 & 77 & $0.384 \pm 0.014$ & $0.811 \pm 0.029$ & 2.11 \\
3 & 67 & $0.444 \pm 0.015$ & $1.102 \pm 0.020$ & 2.48 \\
\rowcolor[HTML]{EFEFEF} 
4 & 109 & $0.612 \pm 0.013$ & $1.433 \pm 0.032$ & 2.34 \\
5 & 120 & $1.096 \pm 0.026$ & $2.089 \pm 0.030$ & 1.90
\end{tabular}
\caption{The interpolation time for the CPR method with Fourier transform and direct interpolation. The first column represents the number $N$ of spins of the Ising model under analysis, the second column represents the number $M$ of spectral elements retained by the procedure described in Sec. and the last column represents the ratio between the direct and CPR interpolation times.}
\label{tab:fft_vs_direct}
\end{table}

\section{Application: two neutrons dynamics} \label{sec:application}

\subsection{Nuclear Theory Background}
We now present the application of the CPR method to two neutron dynamics following Ref. \cite{holland2019optimal}. Nuclear physics is a low-energy theory in which the forces between nucleons are a residual of the color interactions from the underlying quantum chromodynamics theory. In order to find an approximation of the nuclear forces between nucleons, we rely on (chiral) Effective Field Theory (EFT), which works by considering only the interactions determined by energy scales below a certain cutoff scale, while aggregating in the coupling constraints of low energy theory those interactions that affect higher energies \cite{machleidt2011chiralEFT}. The leading order (LO) of the (chiral) EFT expansion captures the features of the neutron interaction, including a tensor-like spin-dependent component, in agreement with experimental measurements \cite{Goodman1980Gamow,Love1981NNinteraction}.In general, single pion exchange is the main interaction mechanism at medium distances ($\approx. 2 e-15 m$), while all other processes (multiple pion exchange or heavier mesons) at shorter distances can be recombined into a spin-dependent contact force.
The LO Hamiltonian ${H_{LO}}$ obtained by EFT is ${H}_{LO}={T}+{V}_{SI} + {V}_{SD}$, where ${T}$ is the kinetic energy, ${V}_{SI}$ is the spin-independent (SI) part of the two-nucleon potential acting on the spatial degrees of freedom of the system, and ${V}_{SD}$ is the spin-dependent (SD) part acting on the spin degrees of freedom.
The SD potential takes into account a vector and a tensor force, namely:
\begin{eqnarray}\label{eq:SD_exact_propagator}
    {V}_{SD}(\mathbf{r})=A^{(1)}(\mathbf{r})\sum_{\alpha}\sigma_{\alpha}^1\sigma_{\alpha}^2+    \sum_{\alpha \beta}\sigma_{\alpha}^1 A^{(2)}_{\alpha \beta}(\mathbf{r})\sigma_{\alpha}^2,
\end{eqnarray}
where $\mathbf{r}$ represents the relative position of the two neutrons in Cartesian coordinates, $\sigma_{\alpha}^k$ for $\alpha=x,y,z$ are the Pauli matrices acting on the spin $k=1,2$, and the functions $A^{(1)}(\mathbf{r})$ and $A_{\alpha \beta}^{(2)}(\mathbf{r})$ are functions derived from the EFT expansion. Their explicit form can be found in Ref.  \cite{gezerlis201cEFT,tews2016neutroncEFT} and in Appendix \ref{sec:appA} the form used in this paper is reported. 

The temporal evolution of a given state  $\ket{\psi(s)}$ is, as always, provided by:
\begin{eqnarray}
\ket{\psi(s)}=e^{-i{H}_{LO}t } \ket{\psi(0)}.
\end{eqnarray}
In the short time limit the dynamics can be approximated, as usual, trotterizing the propagator in small steps of duration $\delta s$:
\begin{eqnarray}
\ket{\psi(s + \delta s)}=e^{-i{H}_{LO}\delta s} \ket{\psi(s)}.
\end{eqnarray}
In this case, the propagator ${U}(\delta s)=e^{-i{H}_{LO} \delta s}$ can be broken up in its two components:

\begin{eqnarray}
&{U}(\delta s)&=[e^{-i({T}+{V}_{SI}) \delta s}] [e^{-i{V}_{SD}(\mathbf{r}) \delta s}] \nonumber \\
&&={U}_{SI}(\delta s){U}_{SD}(\delta s,\mathbf{r}).
\end{eqnarray}

\noindent Always following Ref. \cite{holland2019optimal}, we make use of the approximation that the SD and SI parts can be simulated separately. In this approximation, the SI part only affects the spatial part of the system, while the SD part only acts on the spin degrees of freedom. Given a complete set of states $\ket{r, \varsigma_1, \varsigma_2} \equiv \ket{ r}\otimes \ket{\varsigma_1,\varsigma_2}$, normalized as $\langle {r},\varsigma_1\varsigma_2| {r}',\varsigma_1'\varsigma_2'\rangle=\delta({r}-{r}')\delta_{\varsigma_1'\varsigma_1}\delta_{\varsigma_2'\varsigma_2}$, with $\ket{{r}}$ the (relative) position state, and $\ket{\varsigma_1,\varsigma_2}$ the spin state of the system, we can project the state $\ket{\psi(s)}$ onto this basis at an evolved time $s+\delta s$ as:

\begin{eqnarray}
    && \braket{ r, \varsigma_1, \varsigma_2}{\psi(s+ \delta s)} \simeq \nonumber \\
    && \sum_{\varsigma_1',\varsigma_2'}\int d^3  r' \bra{r}  {U}_{SI}(\delta s) \ket{r'} \times  \\
    && \bra{\varsigma_1,\varsigma_2} {U}_{SD}(\delta s,\mathbf{r}) \ket{\varsigma_1',\varsigma_2'} \braket{r',\varsigma_1',\varsigma_2'}{\psi(s)}\nonumber.
\end{eqnarray}
Therefore, for an infinitesimal time step, one can advance the spatial part and then, with the neutron position fixed, advance the spin part of the wave function.
We exploit this approximation to use a hybrid computing protocol in which the spatial part of the system is simulated with classical algorithms on a classical computer, while the spin dynamics part is performed by a quantum processor. 
This method relies on the saddle point approximation of the path integral of the SI part \cite{smirnov2010saddle}. In this approximation we neglect the quantum fluctuations and compute the classical trajectory of the particle, knowing that it is the most likely one.

In the rest of the paper, we adopt this approach and simulate the trajectory of the two neutrons using a classical differential equation integrator algorithm. At the same time, we advance the spin dynamics by simulating quantum computer results obtained by applying the short-time propagator ${U}_{SD}(\delta s,\mathbf{r})$ to the "instantaneous" spin state at each time step of the spatial trajectory, i.e. corresponding to the "instantaneous" neutrons relative position $\mathbf{r}\equiv \mathbf{r}(s)$. We are interested in the occupation probability of each spin configuration at each time-step, starting from an initial configuration $\ket{\varsigma_1^0,\varsigma_2^0}$, i.e.:
\begin{eqnarray}\label{eq:occupation_prob}
    P_{\varsigma_1,\varsigma_2}(\mathbf{r}(t),t)=|\bra{\varsigma_1,\varsigma_2} {U}_{SD}(\delta t,\mathbf{r}(t)) \ket{\varsigma_1^0,\varsigma_2^0}|^2.
\end{eqnarray}

\subsection{Simulation Setup}

For the scope of this work, we assume that we are working with a single four-level qudit, as in Ref. \cite{holland2019optimal,wu2020high}.
The Hamiltonian of the whole system takes the particular form:
\begin{eqnarray}\label{eq:H_qudit}
H^{qd}=\omega b^{\dagger}b + \frac{\alpha}{2}b^{\dagger}b(b^{\dagger}b-1) + \epsilon(t)(b^{\dagger}+b)
\end{eqnarray}
where we used $\omega=5.114 / 2 \pi \ GHz$ and $\alpha=-0.33 \ GHz$. 
To simplify the optimization procedure, we move to the rotating frame close to the drive frequency $\omega_d$ \cite{holland2019optimal,wu2020high,kirchhoff2018optimized} obtaining the Hamiltonian:

\begin{eqnarray}\label{H_qudit_RWA}
    H^{qd}=&&\Delta b^{\dagger}b + \frac{\alpha} {2}b^{\dagger}b(b^{\dagger}b-1) + \nonumber \\
    &&\epsilon_I(t)(b^{\dagger}+b) +i\epsilon_Q(t)(b-b^{\dagger}),
\end{eqnarray}
where $\Delta =\omega-\omega_d$ is the detuning of the qudit frequency $\omega$ from the drive frequency $\omega_d$ and $\epsilon_I(t)$ and $\epsilon_Q(t)$ are the in-phase and quadrature component of the original control $\epsilon(t)$ in the rotating frame.  $\Delta=0$ since we choose $\omega_d=\omega$.

Now, the first step to run the simulation is to define the mapping between the spin configuration of the neutron system and the quantum processor. 
Two neutrons interacting have four spin configurations: $\ket{\downarrow \downarrow}$,$\ket{\uparrow \downarrow}$,$\ket{\downarrow \uparrow}$, and $\ket{\uparrow \uparrow}$. 
We map each spin configuration to a level of the four-level qubit, namely $\ket{\downarrow \downarrow} \to \ket{0}, \ket{\uparrow \downarrow} \to \ket{1},\ket{\downarrow \uparrow} \to \ket{2},\ket{\uparrow \uparrow}\to \ket{3}$.
The use of the qudit, instead of 2 qubits, is intended to demonstrate how the optimal control protocol is particularly useful in the case of multi-level devices, where the standard gate-based approach is not naturally applicable. Qudits, although technically more difficult to implement, offer the advantage of having more levels to encode information. This makes it possible to reduce the number of standard qubits required and, consequently, the coupling between them, which is a source of noise and error. Qudits are a promising platform for quantum computing \cite{goerz2017charting,neeley2009emulation}. Furthermore, with the analysis in this section, we also want to show how the CPR method can be naturally applied to this type of device.

The second step to run the simulation is to choose a suitable short time interval $\delta s$ and initial conditions for the position $\mathbf{r}_0$ of the neutrons and the spins $\ket{s_0}$. 

Finally, the hybrid classical-quantum simulation procedure, for every time step starting from the initial conditions, is: 
\begin{enumerate}
    \item Update neutrons relative position $\mathbf{r}_{i+1}$ with a classical algorithm on a classical computer.
    \item Evaluate ${U}_{SD}(\delta s,\mathbf{r}_{i+1})$ for the new position $\mathbf{r}_{i+1}$.
    \item Optimize with GRAPE the control pulses $\epsilon_I(t)$ and $\epsilon_Q(t)$ implementing ${U}_{SD}(\delta s,\mathbf{r}_{i+1})$.
    \item Send the controls in the quantum processor to advance the spin state $\ket{\varsigma_{i+1}}$.
\end{enumerate}

This is the theoretical procedure in which steps 1,2 and 3 are performed on a classical computer while step 4 is done on a quantum computer. A schematic representation of this procedure is shown in Fig. \ref{fig:coprocessing_protocols}(a).

\subsection{Simulation with CPR Method}

In the aforementioned procedure, step 3 is the bottleneck of the simulation for the reasons already discussed in Sec.\ref{sec:CPR}. In fact, the optimization algorithm is computationally expensive, depending exponentially on the dimension of the unitary propagator. Moreover, in this application in particular, ${U}_{SD}(\delta s, \mathrm{r})$ depends on the position of the particles, and consequently the corresponding controls must be computed at each time step of the simulation. Furthermore, since many simulations with different initial positions, momenta, and spin states must be performed to fully characterize the dynamics of the system, the number of these controls increases considerably.
To mitigate this problem, in step 3 we replace the GRAPE optimization with the CPR method interpolation.

First, we fix the origin of the coordinate system on one particle. We use the spherical coordinate $(r,\phi,\theta)$ to represent the relative position $\mathbf{r}$ of the second particle. This shifts the dependence of the spin propagator from $\mathbf{r}$ to $(r,\phi,\theta)$. We now need to define the grid $\Lambda_m$ to compute the sets $\mathcal{E}_{z}(t,\Lambda_m)$ where $z=I,Q$ denotes the dataset of $\epsilon_I(t)$ and $\epsilon_Q(t)$ respectively. In this case, $r$ is taken in the interval $I_r=[0.001,2.8]$ fm, where most of the interesting part of the neutron-neutron dynamics takes place, and the interval is discretized in $Z_r=20$ equidistant points. $\phi$ is taken in its natural domain $I_{\phi}=[-\pi/2,\pi/2]$, divided into $Z_{\phi}= 25$ points. Instead $\theta$ takes value in the range $I_{\theta}=[0,\pi/2]$ discretised in $Z_{\theta}=13$ points. The range of $\theta$ is restricted to $[0,\pi/2]$ instead of the usual $[-\pi,\pi]$ because we exploit the symmetries of the quantum system. In fact, the controls of the $[-\pi/2,-\pi]$ interval sector are the same as the $I_{\theta}$ interval, while the controls of the $[0,-\pi/2]$ and $[\pi/2,\pi]$ sectors have the same controls as the $I_{\theta}$ interval, but reversed. So $\Lambda_m$ becomes the three-dimensional grid of all possible combinations $\Lambda'=[r^{a_1},\phi^{a_2},\theta^{a_3}]$ with $a_1\in \{1,. ..,Z_r \}, a_2 \in \{1,...,Z_{\phi}\}$ and $a_3 \in \{1,...,Z_{\theta} \}$.

We compute the datasets $\mathcal{E}_{z}(t,\Lambda_m)$ with GRAPE, imposing a fidelity of $0.9999 \pm 0.00007$. We use parallel computing routines to optimize several controls at once. The first control is computed for $\Lambda'=[r^1,\phi^{13},\theta^1]$ with an all zero initial guess. All other controls are found using the first optimized control as the initial guess. This allows a continuously varying family of controls to be obtained, avoiding solutions falling into different local minima. This makes the interpolation more efficient.

\begin{figure}
    \centering
    \includegraphics[width=0.45\textwidth]{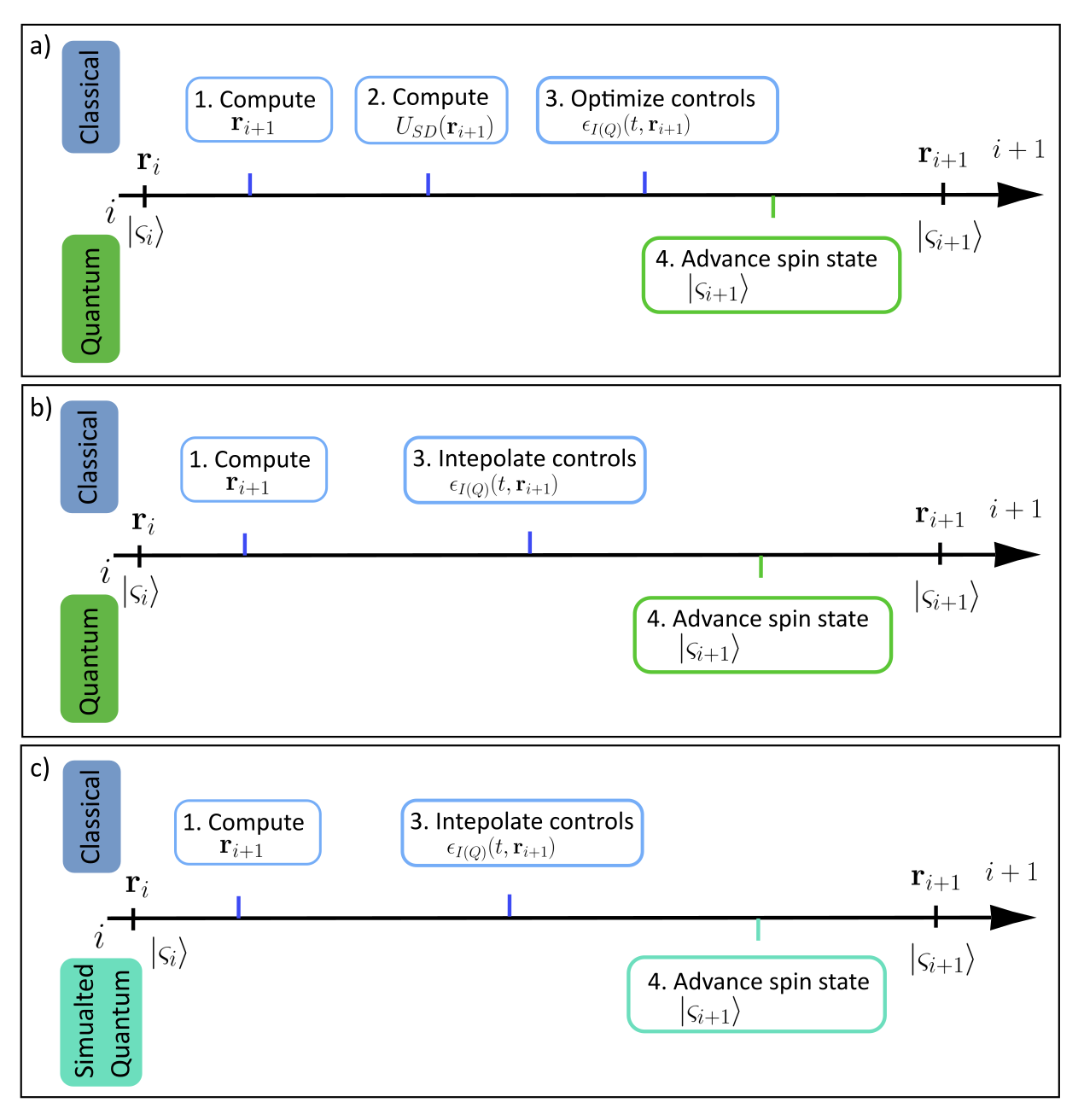}
    \caption{Schematic representation of the hybrid classical-quantum simulation protocols for quantum systems introduced in Sec. \ref{sec:application}. Each panel shows the steps of the algorithm for each time interval $i$ of the quantum simulation. The division of the computations performed on the classical computer (top) and on the quantum computer (bottom) is also shown schematically. Panel (a): Simulation procedure with controls obtained with optimization algorithms.
Panel (b): Simulation procedure with interpolation of controls using the CPR method. Panel (c): Modification of the protocol of panel (b) in which the part carried out on the quantum computer is actually carried out on a simulator. This is the one actually used in this paper and its results are shown in Fig. \ref{fig:simulation}.}
    \label{fig:coprocessing_protocols}
\end{figure}

To summarise, the complete simulation procedure using the CPR method, for each time step starting from the initial conditions, is as follows:

\begin{enumerate}
    \item Update the particles relative position $\mathbf{r}_{i+1}$ with a classical algorithm.
    \item Use the CPR method to interpolate the controls correspondent to new position $\mathbf{r}_{i+1}$ (expressed in terms of the parameters $(r_{i+1},\phi_{i+1}, \theta_{i+1})$).
    \item Obtain ${U}^{rec}_{SD}(\delta s,[r_{i+1},\phi_{i+1}, \theta_{i+1}])$ with Eq. \eqref{eq:propagator_reconstructor} 
    \item Use ${U}^{rec}_{SD}$ to update the spin state as $\ket{\varsigma_{i+1}}=U^{rec}_{SD}\ket{\varsigma_i}$.
\end{enumerate}
This procedure, as in the previous case, assumes that steps 1,2,3 are performed on a classical computer, and step 4 on a quantum computer. A schematic representation of this is given in Fig. \ref{fig:coprocessing_protocols}(b).

However, we test this procedure at a device-simulation level. So once we have reconstructed the appropriate controls $\tilde{\epsilon}_{I(Q)}(t)$, we do not send them into a real quantum computer, but we reconstruct the corresponding ${U}^{rec}_{SD}$ with the appropriate form of the right-hand side of Eq. \eqref{eq:GRAPE} and use it to propagate the spin state as $\ket{\varsigma_{i-1}}={U}_{SD}\ket{\varsigma_i}$. The schematic representation of this modified procedure is reported in Fig. \ref{fig:coprocessing_protocols}(c).

The formula to obtain ${U}^{rec}_{SD}$ in this case is:
\begin{eqnarray}\label{eq:propagator_reconstructor}
&&{U}^{rec}_{SD}(\delta s,[r_i,\phi_i,\theta_i])=\nonumber\\
&&\exp \left \{ -i \delta \tau \sum_{t_i=0}^{N_{\tau}} \frac{\alpha} {2}b^{\dagger}b(b^{\dagger}b-1) \right .  + \tilde{\epsilon}_R(t_i)(b^{\dagger}+b) \nonumber\\
&& \left . + i \tilde{\epsilon}_I(t_i)(b-b^{\dagger}) \right \}.
\end{eqnarray}
Regarding the initial conditions, in this paper we have chosen $\mathbf{r}_0=[r_0^x,r_0^y,r_0^z]=[0.5,-2,0.5] $ fm and $\ket{\varsigma_0}= \ket{0} \equiv \ket{\downarrow\downarrow}$.
The spatial trajectory is obtained by
solving the Newton equations of motion of the particles (i.e. neutrons) with a simple Crank–Nicolson scheme.

\subsection{Results}

\begin{figure*} 
    \centering
    \subfloat[Neutrons trajectory]{%
        \includegraphics[width=0.4\textwidth]{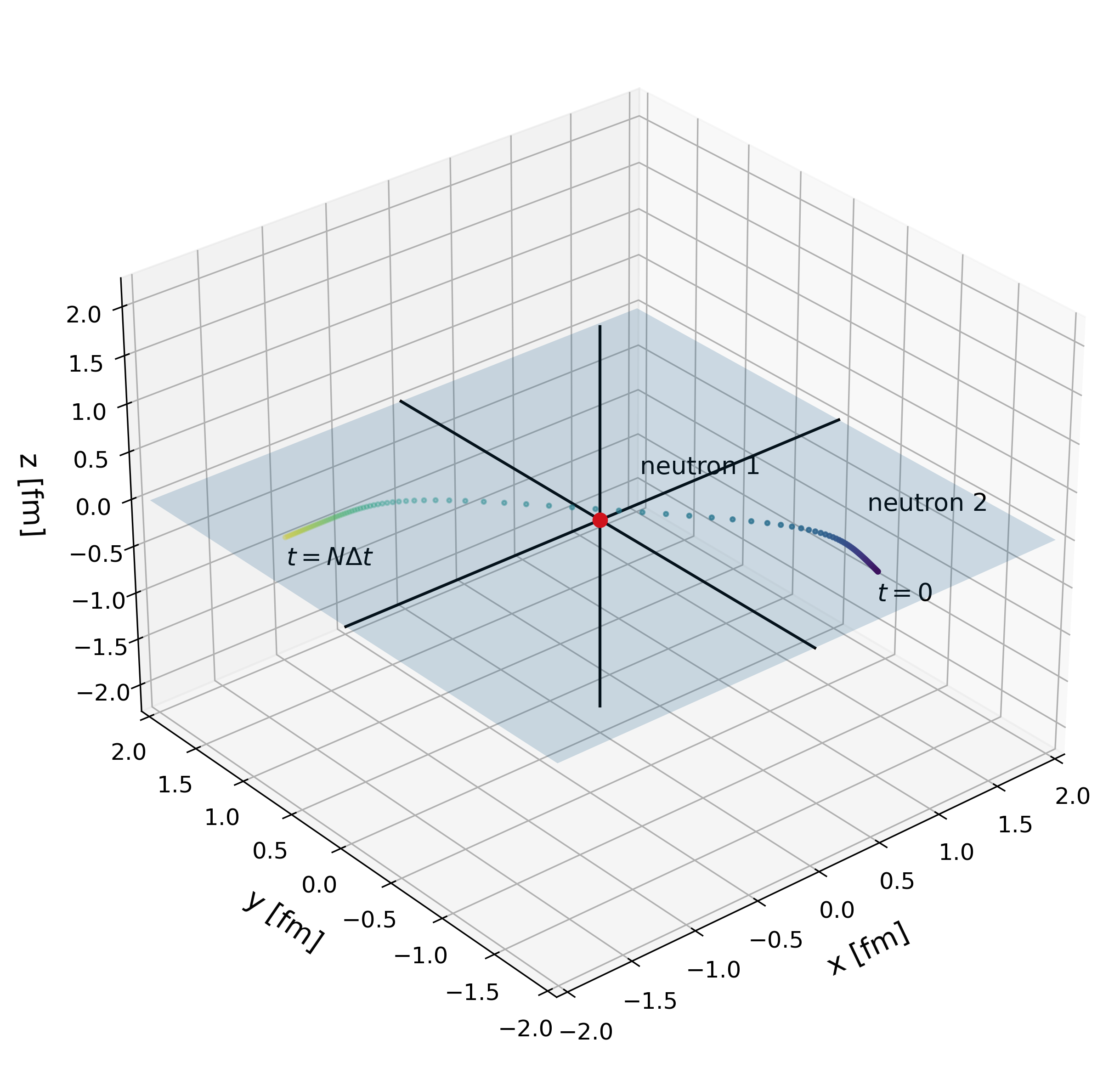}%
        \label{fig:a}%
        }%
    \hfill%
    \subfloat[Spin dynamics]{%
        \includegraphics[width=0.5\textwidth]{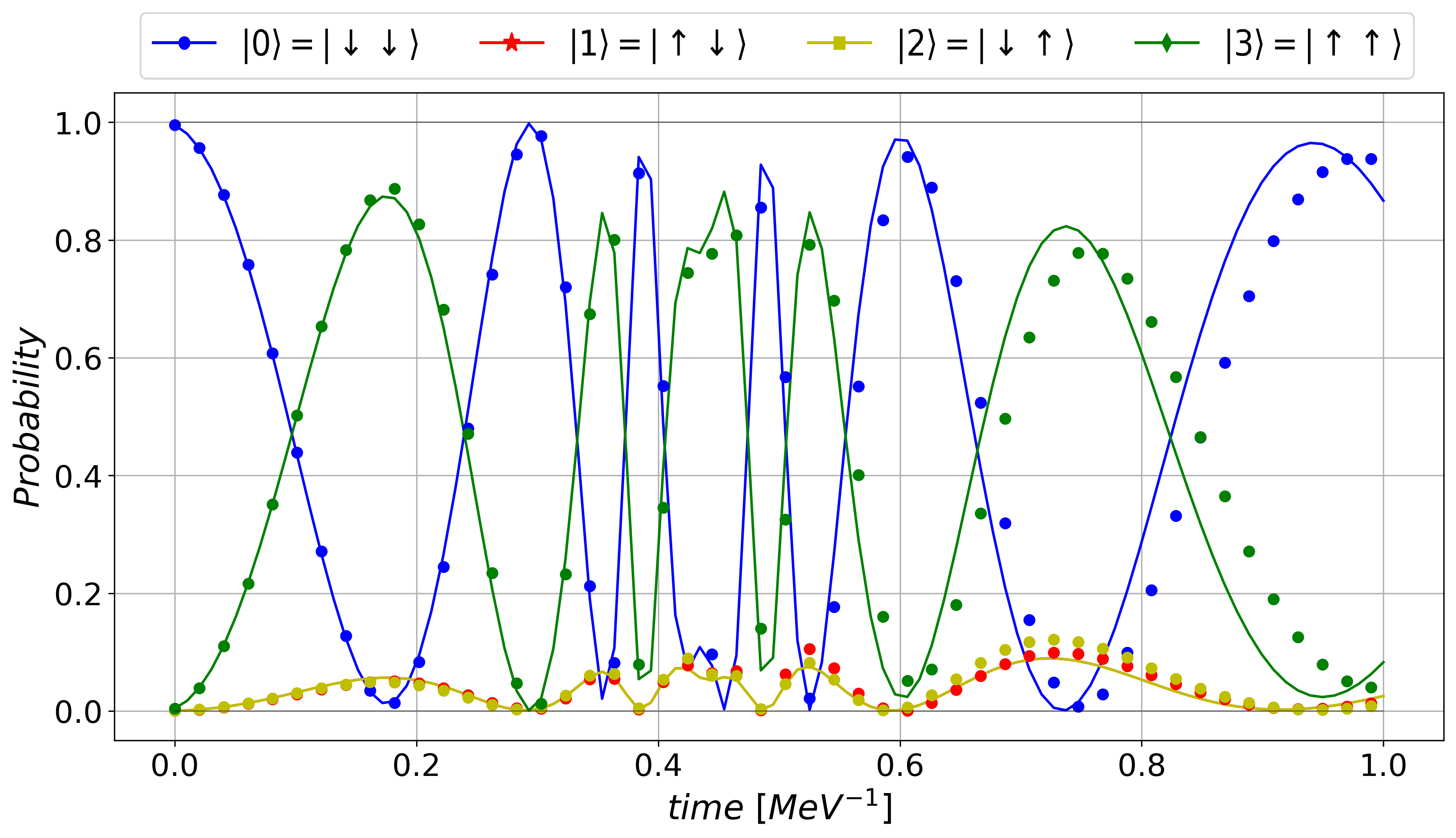}%
        \label{fig:b}%
        }%
    \caption{Representation of the neutrons dynamics. \textit{Panel (a)}: A single realization of a classical spatial trajectory for two neutrons obtained by solving their equation of motion with a Crank-Nicolson scheme starting from a specified initial condition. The origin of axes is fixed on one particle. \textit{Panel (b)}: Spin dynamics in terms of occupation probability corresponding to the trajectory of the panel (a). For every time $s$ the value is found using Eq. \eqref{eq:occupation_prob}. }\label{fig:simulation}
\end{figure*}
We carry out the simulation, following the procedure just introduced, for 100 time steps. The results of this simulation are shown in Fig. \ref{fig:simulation}. Panel (a) shows the trajectory of the second particle with respect to the first (on which the coordinates are fixed). The colors represent the time of the quantum system. Panel (b) reports the spin dynamics in terms of the occupation probability, computed with Eq. \eqref{eq:occupation_prob}, along the spatial trajectory shown in panel (a). The solid lines represent the exact spin dynamics found by numerically solving the Schr\"{o}dinger equation with the exact time-dependent Hamiltonian, instead, the dots represent the spin state for subsequent applications of ${U}^{rec}_{SD}(\delta s,[r_i,\phi_i, \theta_i])$ starting from the initial state $\ket{\varsigma_0}$ with the controls obtained by the CPR method. It, therefore, represents the dynamics that we would obtain using the controls in the quantum processor (without the noise of the real device). As can be seen, the dotted spin dynamics follow the exact reference dynamics very well, except for an accumulation of errors at the end of the simulation due to the imprecision of the CPR interpolation. The average fidelity between the reconstructed and the exact propagators is $0.9997\pm 0.0005$. 

In general, the accuracy of the method can be modulated using different configurations of the CPR method elements. A finer discretization of the intervals results in higher accuracy. This discretization can also be non-constant. It can be finer only in correspondence of points where the Hamiltonian changes most rapidly with respect to its parameters, and coarser otherwise. In addition, as in the case just presented, one can exploit the symmetries of the quantum system under study to greatly reduce the number of controls to be computed beforehand. In general, then, one can find a good trade-off between accuracy and computational cost.

As mentioned earlier, the main advantage of the CPR method is the shorter time required to derive the controls necessary to implement a unitary transformation. We report the data from the simulation. We use 800 time-step controls (i.e. controls with a duration of 50 ns and a signal sampling frequency of 16 GHz). The GRAPE computation time for this case is $t_G=5.12$ s. The total number $A$ of controls in the data sets $\mathcal{E}_{I(Q)}(t,\Lambda_m)$ is $Z_{r}Z_{\phi}Z_{\theta}=6500$. The total computation time for such a data set, using parallel computing on a mid-range computer with a 2.9 GHz CPU, 4 cores and 16 GM of RAM, is $T_G=10400$ seconds. This means that each control took an average of 1.6 seconds instead of the 5.12 seconds it would have taken using serial computation. This means that by taking advantage of parallel computing and using the same initial condition for each control optimization, we can reduce this time on average, making the CPR method even more efficient. With more CPU cores, this would be even more efficient. The time to compute a single step of the simulation with GRAPE in this setup, for controls of 800 time-steps and a fidelity of $0.9999$, is $6.43 \pm 0.02$ seconds. CPR method shrinks this time to $0.913\pm 0.002$ seconds. Therefore, from a computational point of view, we gain about an order of magnitude by using the CPR method (for this specific configuration). However, the CPR method becomes preferable to GRAPE optimization if the time $T_G$ to compute the entire data set in advance plus the time $t_CPR$ to compute the $P$ controls needed in the simulation is less than the time to compute the same $P$ controls using GRAPE. Formally, $P$ should satisfy $T_G+t_{CPR}P<t_GP$. This is true if $P>T_G/(t_G-t_{CPR})\approx 2396$. So, in this configuration, we get a net benefit from using the CPR method if the number of controls needed during the simulation is greater than 2396. This number is not prohibitive, since in order to characterize the dynamics of the system, or to calculate quantities such as the cross-section, it is necessary to calculate many trajectories with different initial conditions of position and spin so that this limit is easily reached.

\section{Conclusion}
\label{sec:conclusion}
We introduced a general scheme, the Control Pulse Reconstruction (CPR) method, which can be used to improve the performance of quantum computations based on optimal control techniques whenever a frequent evaluation of the control pulses is required. In general, this corresponds to considering a family of Hamiltonians or propagators that depend on some parameters, which in turn are a function of the evolution time. 

The control pulses for the single multi-level gate required to encode a quantum system propagator on the quantum device were obtained by numerical optimization using the GRAPE algorithm as implemented in the QuTiP Python package. By computing in advance a set of control pulses of propagators with different parameter values, then interpolating between their corresponding spectra and Fourier transforming back the interpolated spectrum, it was possible to derive a mathematical model that allows to reconstruct the control pulses of any value of the parameter, bypassing the numerical optimization. 

To investigate the performance and scaling of our approach, we computed the fidelity between the exact and the reconstructed propagator of a transverse-field Ising model for different number of parameters, number of qudits involved and for different parameter grid spacings, obtaining excellent results in all cases. 

As a test of this protocol, we presented the simulation of the real-time evolution of two neutron dynamics with a realistic potential obtained by effective field theory. Their spin configurations were mapped into a controllable four-level trasmon qudit. We have shown that the time required to advance a single time step of the simulation with the CPR method is on average one order of magnitude less than in the GRAPE case (at the device simulation level).
Complex systems, whose Hamiltonian depends on some external parameters, require the control pulses implementing their propagators to be optimized at each time step. The CPR method, which avoids the large amount of time and computational resources required for computation, could be of great interest for improving the simulation of realistic quantum systems. In particular, it opens the possibility of a manageable implementation of several classical-quantum hybrid protocols on multi-level qudits.

\section{Acknowledgment}
This research was partially supported by Q@TN grants ML-QForge (PL) and ANuPC-QS (FT). \newline This work was prepared in part by LLNL under Contract DE-AC52-07NA27344  with support from the Laboratory Directed Research and Development grant 19-DR-005.


\appendix
\section{APPENDIX: Neutrons LO Interaction}\label{sec:appA}

We refer to Ref. \cite{tews2016neutroncEFT} to obtain the explicit form of $A^{(1)}(\mathbf{r})$ and $A^{(2)}_{\alpha \beta}(\mathbf{r})$ of the SD neutron-neutron interaction at LO of the chiral EFT in the coordinate space given by Eq.\eqref{eq:SD_exact_propagator}.
They are:
\begin{eqnarray}
&& A^{(1)}(\mathbf{r})=C_1 \delta_{R_0}(\mathbf{r})-Y_{\pi}(r)(1-e^{-(r/R_0)^4}),\\
&& A^{(2)}_{\alpha \beta}(\mathbf{r})=T_{\pi}(r)(3\frac{r_{\alpha}r_{\beta}}{r^2}-\delta_{\alpha\beta})(1-e^{-(r/R_0)^4} ),
\end{eqnarray}
The SI part instead can be written as ${V}_{SI}=C_0\delta_{R_0}(\mathbf{r})$.
In all these expressions, $C_0$ and $C_1$ are experimental constants fit to reproduce some quantity (e.g. the $s$-wave nucleon-nucleon phase shifts), 
\begin{eqnarray}
\delta_{R_0}(\mathbf{r})=\frac{1}{\pi \Gamma(3/4)R_0^3}\exp{(-r/R_0)}
\end{eqnarray} 
is the regulated Dirac function, $Y_{\pi}(r)$ is the Yukawa function, i.e.:
\begin{eqnarray}
Y_{\pi}(r)=\frac{m_{\pi}^3}{12\pi}\left(\frac{g_a}{2f_\pi} \right)^2 \frac{\exp(-m_{\pi}r)}{m_{\pi}r},
\end{eqnarray}
and 
\begin{eqnarray}
T_{\pi}(r)=\left( 1+\frac{3}{m_\pi r}+\frac{3}{m_\pi^2 r^2} \right)Y_\pi(r)
\end{eqnarray}
where $g_a$, $f_\pi$ and $m_{\pi}$ are the axial-vector coupling constant, the pion exchange decay constant and the pion mass.
%
%

%
\bibliographystyle{unsrt}
\bibliography{bibliography.bib}

\end{document}